\documentclass[12pt]{article}
\usepackage[mathscr]{eucal}
\usepackage{amssymb,latexsym}
\usepackage{verbatim}
\usepackage{amsmath}
\usepackage{amsthm}
\usepackage{enumerate}
\usepackage{authblk}
\usepackage{color}
\usepackage{url}

\usepackage{hyperref}

\numberwithin{equation}{section}

\newtheorem{thm}{Theorem}[section]
\newtheorem{prop}[thm]{Proposition}

\newcommand\calT{{\mathcal{T}}}

\renewcommand\l{\lambda}

\renewcommand\S{\Sigma}

\renewcommand\d{\partial}

\newcommand\f{\phi}

\newcommand\D{\nabla}
\newcommand\e{\epsilon}
\renewcommand\b{\beta}

\renewcommand\div{{\rm div}}

\renewcommand\l{\lambda}
\newcommand\g{\gamma}

\renewcommand\th{\theta}

\newcommand\beq{\begin{equation}}
\newcommand\eeq{\end{equation}}
\newcommand\ben{\begin{enumerate}}
\newcommand\een{\end{enumerate}}
\newcommand\bit{\begin{itemize}}
\newcommand\eit{\end{itemize}}

\DeclareMathOperator{\diver}{div}

\renewcommand{\div}{\diver}

\newcommand{\tr}{\mathrm{tr}\,}

\newcommand{\ov}{\overline}

\newcounter{mnotecount}

\title{Remarks on the size of apparent horizons}

\author{Gregory J. Galloway}

\affil{Department of Mathematics
\\University of Miami}

\begin{document}
\date{}
\maketitle  
\vspace{.2in}

\begin{abstract}  
Marginally outer trapped surfaces (also referred to as apparent horizons) that are stable in $3$-dimensional initial data sets obeying the dominant energy condition strictly are known to satisfy an area bound. The main purpose of this note is to show (in several ways) that such surfaces also satisfy a diameter bound.
\end{abstract}


\section{Introduction}
\label{intro}

As reviewed in Section 2, marginally outer trapped surfaces 
(a.k.a.\! apparent horizons) 
admit an important notion of stability, analogous, at some level, to the stability of minimal surfaces in Riemannian manifolds.  Heuristically, a stable marginally outer trapped surface  is {\it infinitesimally} outermost.  It is known that stable marginally outer trapped surfaces in $3$-dimensional initial data sets that obey the dominant energy condition strictly
satisfy an area bound, as we now briefly recall. 

An {\it initial data set} $(M,g,K)$ consists of a smooth orientable $n$-dimensional manifold $M$ equipped with a 
Riemannian metric $g$ and a symmetric $(0,2)$-tensor $K$. 
The main physical example is when $(M,g,K)$ is an initial data set in a spacetime (time oriented Lorentzian manifold) $(\overline{M}, \overline{g})$, i.e. $M$ is a spacelike hypersurface in 
$\overline{M}$, with induced metric $g$ and second fundamental form $K$.  

The {\it local energy density} $\mu$ and the {\it local current density} $J$ of an initial data set $(M,g,K)$ are given by
\begin{align*}
\mu=\frac{1}{2}(S-|K|^2+(\tr K)^2)\quad\mbox{and}\quad J=\div(K-(\tr K)g),
\end{align*}
where $S$ is the scalar curvature of $(M,g)$. When $(M,g,K)$ is a spacetime initial data set, these quantities are given by $\mu = G(u,u)$,  $J = G(u,\cdot)$, where $G$ is the Einstein tensor 
$G = {\rm Ric}_{\overline M} - \frac12 R_{\overline M} \overline g$.

The following was obtained  with A. Mendes; see 
\cite[Proposition 3.1]{GalMen}.  Henceforth we use the shorthand `MOTS' for `marginally outer trapped surface'.

\begin{prop}\label{size1}
Let   $\S$ be a stable
 MOTS in a $3$-dimensional initial data $(M,g,K)$.  Suppose there exists $c > 0$, such that  $\mu +J(\nu) \ge c$ on $\S$, where $\nu$ is the outward unit normal to $\S$.   Then $\S$ is topologically a $2$-sphere and its  area  satisfies,
\beq\label{ineq1}
A(\S) \le \frac{4\pi}{c} \,.
\eeq
Moreover, if equality holds, $\S$ is a round $2$-sphere, with  Gaussian curvature $\kappa_\S = c$, the outward null second fundamental form 
$\chi$  of $\S$ vanishes, and $\mu +J(\nu) = c$ on 
$\S$.
\end{prop}  

In the statement of Proposition 3.1 in \cite{GalMen}, for convenience  it was assumed that $\S$ is topologically a $2$-sphere, but it is immediate from the proof that this is necessarily the case.  

As discussed in~\cite{GalMen}, the equality case holds in certain well known models.  The rigidity part of Proposition \ref{size1} was an important element in proving local and global initial data rigidity results for outer area minimizing marginally outer trapped $2$-spheres; see \cite[Theorem 3.2]{GalMen} and \cite[Theorem 3.1]{GalMen2}.  These results extend to general initial data sets certain results of Bray, Brendle and Neves \cite{BBN} concerning area minimizing $2$-spheres in Riemannian $3$-manifolds with positive scalar curvature.  

We note that a  positive lower bound on the quantity $\mu +J(\nu)$   
arises naturally due to the presence of matter fields and/or a positive cosmological constant.  Suppose  $(M,g,K)$ is an initial data set in a spacetime $(\ov M, \ov g)$ that satisfies the Einstein equation,
\beq
G + \Lambda \bar g = \cal{T}  \,,
\eeq
where $\cal T$ is the energy-momentum tensor.
Setting $\ell = u+\nu$, 
where $\nu$ is any unit vector tangent to $M$  and 
where $u$ is the future directed unit normal to $M$, we have, 
$$
\mu + J(\nu)  = G(u,\ell) = \calT(u,\ell) + \Lambda  \ge \Lambda \,.
$$
provided $\cal T$ obeys the dominant energy condition (which includes the matter vacuum case  $\mathcal{T} = 0$). Hence, for positive $\Lambda$, the inequality \eqref{ineq1} becomes,
$$
A(\S) \le \frac{4\pi}{\Lambda} \,.
$$
Such an inequality was obtained by Hayward, Shiromizu, and Nakao \cite{Hay1} for spacetimes with positive cosmological constant, using a slightly different notion of stability in the spacetime setting.  
Finally,  note that the condition $\mu +J(\nu) \ge c$ is implied by the strict dominant energy condition $\mu - |J| \ge c$.

The main purpose of this note is to point out that stable MOTS in a $3$-dimensional initial data set obey a {\it diameter} bound, as well.  

\begin{thm}\label{size2}
Let   $\S$ be a stable MOTS in a $3$-dimensional initial data set $(M,g,K)$.  Suppose there exists $c > 0$, such that  $\mu +J(\nu) \ge c$ on $\S$, where $\nu$ is the outward unit normal to $\S$.   Then the diameter of $\S$ satisfies,
\beq\label{ineqdiam}
{\rm diam}(\S) \le  \frac{2}{\sqrt {3}}\cdot \frac{\pi}{\sqrt{c}} \, .     
\eeq
\end{thm}

Here  ${\rm diam}(\S)$ has the usual meaning,
$$
{\rm diam}(\S)  = \sup_{x, y \in \Sigma} d(x,y) \,,
$$
where $d(x,y)$ is the distance between $x$ and $y$ in $\S$ with respect to the induced metric on~$\S$.  The obvious advantage of a diameter estimate, is that when $\mu +J(\nu)$ is large, $\S$ must, roughly speaking, be small in all directions. In general, surfaces with small area can still have very large diameter.  We note, however, whereas the bound in \eqref{ineq1} is sharp, the bound in \eqref{ineqdiam}  is not.  

As we shall discuss in Section 3, Theorem \ref{size2} can be obtained from one of the recent {\it spectral torical band inequalities} for $n$-dimensional manifolds, $2 \le n \le 7$, by Hirsch et al., \cite[Theorem 1.3]{HKKZ}, whose proof is based on the theory of $\mu$-bubbles, together with a key property of stable MOTS \cite{GalSch} (see Section 2). 
In Section 2, we also present a direct proof of Theorem \ref{size2}, using only elementary techniques, in addition to  this property of MOTS.  Some comments about higher dimensions are presented at the end of Section 3.
In what follows all manifolds are assumed to be smooth, connected and orientable, unless otherwise stated.

\section{Preliminaries and a proof of Theorem \ref{size2}}

\subsection{Preliminaries}  Let $(M,g,K)$ be an $n$-dimensional, $n \ge 3$, initial data set.  One can always arrange for $(M,g,K)$ to be an initial data set in a spacetime  
$(\overline{M}, \overline{g})$ (see e.g.\! \cite[Section 3.2]{AM2}), and we find it convenient to do so. However, all essential quantities depend only on the initial data.   Thus we may assume $M$ is a spacelike hypersurface, with induced metric $g$ and second fundamental form $K$, in an $(n+1)$-dimensional spacetime
$(\overline{M}, \overline{g})$. Our sign convention is such that, for vectors $X,Y \in T_pM$, $K$ is defined as, $K(X,Y) = \ov g(\bar \D_X u,Y)$, 
where $u$ is the future directed timelike unit normal  field to~$M$. 

Let $\S$ be a closed  two-sided surface in $M$.   Then $\S$ admits a smooth unit normal field $\nu$ in $M$, unique up to sign.  By convention, we refer to such a choice as outward pointing. 
Then $l_+ = u+\nu$ and $l_- =  u - \nu$ are future directed outward pointing and inward pointing, respectively,  null normal vector fields along $\S$.  
The null second fundamental forms $\chi_+$ and $\chi_-$ of $\Sigma$ in $\bar M$ are defined by 
\beq\label{nullforms}
\chi_\pm(X,Y)=\ov g(\bar\nabla_Xl_\pm,Y) ,
\quad X,Y\in T_p\Sigma \,.
\eeq
The null expansion scalars $\theta_\pm$ of $\Sigma$ are obtained by tracing 
$\chi_\pm$, 
\beq
\theta_\pm=\tr_\Sigma\chi_\pm=  \div_{\S} \ell_{\pm}  \,.
\eeq 
Physically, $\th_+$ measures the divergence of the outgoing light rays from $\S$.
In terms of the initial data $(M, g, K)$,
$$
\th_{\pm}  = {\rm tr}_{\S} K \pm H \,,
$$
where $H = \div_{\S} \nu$ is the mean curvature of $\S$ within $M$.  

Penrose famously defined $\S$  to be a trapped surface if both $\th_-$ and $\th_+$ are negative.   Focusing attention on the outward null normal, we say that  
$\S$ is an outer trapped surface  if $\th_+ < 0$.  If $\th_+$ vanishes identically we say  that $\S$ is a marginally outer trapped surface (or MOTS).    Note that in the 
`time-symmetric case' ($K= 0$) a MOTS is simply a minimal surface.

We now recall the notion of stability of MOTS as introduced by Andersson, Mars and Simon (\cite{AMS1, AMS2}; see also \cite{Lee}).

Let $\S$ be a MOTS in the initial data set $(M,g,K)$ with outward unit 
normal~$\nu$.  We consider a normal variation of $\S$ in $M$,  i.e.,  a variation 
$t \to \S_t$ of $\S = \S_0$ with variation vector field 
$V = \frac{\d}{\d t}|_{t=0} = \phi\nu$,  $\phi \in C^{\infty}(\S)$.
Let $\th_+(t)$ be the null expansion of $\S_t$
with respect to $l_t = u + \nu_t$, where 
$\nu_t$ is the
outer unit normal  to $\S_t$ in $M$.   A computation gives,
\beq\label{thder} 
\left . \frac{\d\th_+}{\d t} \right |_{t=0}   =
L(\f) \;, 
\eeq 
where $L : C^{\infty}(\S) \to C^{\infty}(\S)$ is the operator, 
\begin{align}\label{stabop}
L(\phi)  &= -\triangle \phi + 2g(X,\D\phi)  + \left(\mathcal{Q} +{\rm div}\, X - |X|^2 \right)\phi \, , \\
\mathcal{Q} &=  \frac12 S_{\S} - (\mu + J(\nu)) - \frac12 |\chi_+|^2\,  ,
\end{align}
where $\triangle$, $\D$ and ${\rm
div}$ are the Laplacian, gradient and divergence operators,
respectively, on $\S$, $S_{\S}$ is the scalar curvature of $\S$ 
and $X$ is the 
vector field  on $\S$  dual to the one form $K(\nu,\cdot)|_{T\S}$. 

The operator $L$ is not self-adjoint in general, but nevertheless admits a principal eigenvalue.
\begin{prop}\label{prin}  There is a real eigenvalue $\l_1 = \l_1(L)$, called the principal eigenvalue, which is an eigenvalue with smallest real part.
The associated eigenfunction $\f$, $L(\f) = \l_1 \f$, is unique up to a multiplicative constant, and can be chosen to be strictly positive.
\end{prop}

In the time-symmetric 
case ($K =0$), $L$ reduces to  the classical stability (or Jacobi) operator for minimal surfaces in Riemannian manifolds.  As such, we refer to $L$ as the MOTS stability operator.  
In analogy with the minimal surface case, we
say that a MOTS is stable provided $\l_1(L) \ge 0$.  (In the minimal surface case this is equivalent to the second variation of area being nonnegative.)
Note that, by taking $\phi > 0$ to be the principal eigenfunction in \eqref{thder}, a stable MOTS $\S$ admits an outward variation $t \to \S_t$  with $\frac{\d\th_+}{\d t}|_{t=0} \ge 0$.  It can be shown, in a certain sense, that the converse also holds.
 
Stable MOTSs arise in various
situations. For example, {\it weakly outermost} MOTSs are stable.  A separating MOTS 
$\S$ is weakly outermost in $(M,g,K)$ if there are no outer trapped surfaces outside of and homologous to $\S$. Indeed, if $\l_1(L) < 0$,  \eqref{thder} implies that $\S$ can be deformed outward to an outer trapped surface.  Weakly outermost MOTSs include, in particular, compact cross sections of the event horizon in stationary black hole spacetimes obeying the null energy condition, as well as the MOTS foliating a dynamical horizon; cf. \cite{AshGal}.  More generally, the boundary of the trapped region is a weakly outermost MOTS (see e.g. \cite{AM2, Eich2}).
  
The proof of Theorem \ref{size2} in Section 2 relies on the following key fact. Consider the 
``symmetrized'' operator
$\hat{L}: C^{\infty}(\S) \to C^{\infty}(\S)$,
\beq\label{symop}
\hat{L}(\phi)  = -\triangle \phi  + \mathcal{Q}  \phi \,,
\eeq
obtained formally from \eqref{stabop}  by  setting $X= 0$. 

\begin{prop}[\cite{GalSch}]\label{keyfact} If $\S$ is a stable MOTS in an $n$-dimensional, $n\ge 3$,
initial data set, then $\S$ is `symmetric stable', i.e.\! $\l_1(\hat{L}) \ge 0$, where 
$\l_1(\hat{L})$ is the principal eigenvalue of $\hat{L}$.
\end{prop}

In fact, the arguments in \cite{GalSch} show that $\l_1(\hat{L}) \ge\l_1(L)$.  These arguments were motivated in part by computations made by Schoen and Yau \cite{SY2} on graphical solutions of Jang's equation in the presence of a translational symmetry, in their proof of the positive mass theorem in the general non-time-symmetric case.

\medskip
\noindent
{\it Remark.} The arguments in \cite{GalSch} apply equally well to the following.  Let $\S^{n-1}$ be a stable MOTS in an $n$-dimensional, $n\ge 3$, initial data set, and let $\Omega$ be a smooth connected compact $(n - 1$)-dimensional
submanifold-with-boundary in $\S$.  
Then $\l_1(\Omega) \ge 0$, where $\l_1(\Omega)$ is the principal Dirichlet eigenvalue of the symmetrized operator $\hat{L}$ restricted to $\Omega$.  This mild variation of Proposition \ref{keyfact}
will be used  in Section 3.

\subsection{Proof of Theorem \ref{size2}}

We now present a proof of the diameter estimate \eqref{ineqdiam}.  
The proof is based 
largely on the proof of Theorem 3.2 in \cite{GalOMar}. 
The latter result was motivated by, and extends to the fully non-time-symmetric case, Theorem 1 in \cite{SYsize}, but uses a different measure of size, one which is amenable to the present situation.

\proof  Fix points $p,q \in \S$ such that ${\rm diam}(\S) = d(p,q)$. 
$\S$ is a stable MOTS, hence by Proposition \ref{keyfact},  
it is symmetric stable, $\l_1 = \l_1(\hat{L}) \ge 0$. Let $\psi > 0$ be an associated eigenfunction.
Substituting
$\phi = \psi$ into Equation \eqref{symop}, we obtain,
\beq\label{laplace}
\triangle \psi = - (\mu + J(\nu)  +  \frac12 |\chi_+|^2 + \l_1 - \kappa) \psi
\eeq
where $\kappa= \frac12 S_{\S}$ is the Gaussian curvature of $\S$ in the induced metric $h$.

Now consider $\S$ in the conformally related metric $\tilde h = \psi^2 h$.   The Gaussian
curvature of $(\S, \tilde h)$ is given by,
\beq\label{relate}
\tilde \kappa = \psi^{-2} \kappa - \psi^{-3} \triangle \psi + \psi^{-4} |\D\psi|^2 \,.
\eeq
Substituting \eqref{laplace} into \eqref{relate} gives,
\beq\label{gauss}
\tilde \kappa = \psi^{-2}(P + \psi^{-2} |\D\psi|^2)   \,,
\eeq
where, 
\beq\label{P}
P = \mu + J(\nu) + \frac12 |\chi_+|^2 + \l_1  \,.
\eeq 

 Let $\g$ be a minimal geodesic from $p$ to $q$ in the metric $\tilde{h}$.   Then  by Synge's formula~\cite{ON}
for the second variation of arc length, we have along $\g$,
\beq\label{ineq}
\int_0^{\tilde \ell} \left(\frac{df}{d\tilde s}\right)^2 - \tilde \kappa f^2\, d \tilde s \ge 0 \,,
\eeq
for all smooth functions $f$ defined on $[0,\tilde \ell]$ that vanish at the end points, where
$\tilde \ell$ is the $\tilde h$-length of $\g$ and $\tilde s$ is $\tilde h$-arc length along $\g$.
By making the change of variable $s  = s(\tilde s)$, where $s$ is $h$-arc length along $\g$,
and using Equation \eqref{gauss}, \eqref{ineq} becomes,
\beq\label{ineq2}
\int_0^{\ell} \psi^{-1}(f')^2 - (P + \psi^{-2} |\D\psi|^2)\psi^{-1}  f^2  \, d s \ge 0 \,,
\eeq
for all smooth functions $f$ defined on $[0,\ell]$ that vanish at the endpoints, where
$\ell$ is the $h$-length of $\g$, and $' = \frac{d}{ds}$.

Setting $k= \psi^{-1/2}f$ in \eqref{ineq2},    we obtain by a small
computation,
\beq\label{ineq3}
\int_0^{\ell}  (k')^2  -  P\,k^2 -  \big( \frac34\psi^{-2}(\psi')^2k^2 - \psi^{-1}\psi'kk'\big) \, ds \ge 0  \,,
\eeq
where $\psi'$ is shorthand for $(\psi \circ \g)'$, etc.
Completing the square on the bracketed term of the integrand shows that,
\beq
\frac34\psi^{-2}(\psi')^2k^2 - \psi^{-1}\psi'kk' \ge - \frac13(k')^2 \,,  \nonumber
\eeq
which then  implies,
\beq\label{ineq4}
\int_0^{\ell}  \frac43(k')^2  -  P\,k^2 \, ds \ge 0 \,.
\eeq
Since,  from \eqref{P},  $P \ge \mu + J(\nu) \ge c$,
\eqref{ineq4}  gives,
\beq\label{ineq5} 
\frac43\int_0^{\ell}  (k')^2 \,ds \ge  c  \int_0^{\ell} k^2 \, ds \,.
\eeq
Setting $k = \sin \frac{\pi s}{\ell}$  in \eqref{ineq5} then gives,
\beq
\ell \le  \frac{2}{\sqrt {3}}\cdot \frac{\pi}{\sqrt{c}}\,. \nonumber
\eeq 
Since ${\rm diam}(\S) \le \ell$, the diameter bound \eqref{ineqdiam} is 
established.\qed
\newpage

\noindent
{\it Remark.}  From the inequality $\mu + J(\nu) \ge c$,
and the fact that $\l_1(\hat{L}) \ge 0$, 
one can also derive from \eqref{laplace} that the principal eigenvalue of the operator  $-\triangle \phi + \kappa\, \phi$ is greater than or equal to $c$, i.e., 
$\l_1(-\triangle \phi + \kappa\, \phi) \ge c$.   In the case of  $2$-dimensional  compact (or, more generally, complete) Riemannian manifolds, Kai Xu has shown for $\b > \frac14$, that the eigenvalue condition $\l_1(-\triangle \phi +\b \kappa\, \phi) \ge c$ leads to a diameter bound depending on $\b$; see \cite[Theorem 1.4]{Xu}.  In the case $\b = 1$, one obtains the diameter bound \eqref{ineqdiam}.  Xu gives two proofs, one which uses a method similar to that used here,  and one which uses the theory of $\mu$-bubbles.

%

\section{A spectral bound inequality of Hirsch et al. 
\cite{HKKZ}
}

Given a smooth $n$-dimensional Riemannian manifold $(M^n,g)$ and real number $\xi$, in \cite{HKKZ} the authors define what they call 
the {\it $\xi$-spectral constant}  $\Lambda_{\xi}$.  As they point out,  when $(M^n,g)$ is a compact manifold with boundary, $\Lambda_{\xi}$ is the principal Dirichlet eigenvalue 
of the operator
\beq
L_{\xi}(\phi) =  -\triangle \phi  + \xi R \phi \, ,
\eeq
where $R$ is the scalar curvature of $(M^n,g)$.

A compact Riemannian manifold-with-boundary $(N^n, g)$ whose boundary components are separated into two disjoint and non-empty collections, 
$\d N^n=\d_- N^n\sqcup \d_+ N^n$, is referred to as a Riemannian band.  Following \cite{HKKZ}, a Riemannian band is called \textit{overtorical} if there exists a smooth map\footnote{In some places in the literature, the band map $F$ is only required to be continuous. In fact, while not needed here, Theorem 1.3 in \cite{HKKZ} still holds in this case.  There are situations where this may be useful. We thank Marcus Khuri for a communication regarding this point.}
$F:N^n\to T^{n-1}\times[-1,1]$ of nonzero degree, with $F(\d_\pm N^n)\subset T^{n-1}\times \{\pm1\}$.

The following result, which assumes $\xi =\frac12$, is proved in \cite{HKKZ}.  
\begin{thm}[\cite{HKKZ}, Theorem 1.3]\label{mu bubble}
Let $(N^n,\d_{\pm}N^n,g)$ be an  overtorical band with $2 \le n\leq 7$. If $\Lambda_{\frac{1}{2}}>0$ then
\beq\label{distance}
d(\d_- N^n,\d_+ N^n)\le\pi\sqrt{\frac{2n}{(n+1)\Lambda_{\frac{1}{2}}}} \,,
\eeq
where $d(\d_- N^n,\partial_+ N^n)$ is the distance between $\d_- N^n$ and $\d_+ N^n$.
\end{thm}

\noindent
We refer to $d(\d_- N^n, \d_+ N^n)$ as the {\it width} of the overtorical band. 

\smallskip
As noted in the introduction, the proof of Theorem \ref{mu bubble} is based on the theory of $\mu$-bubbles.  The dimension restriction, $n \le 7$, comes from the regularity theory of $\mu$-bubbles, which is tied to the regularity theory of surfaces of prescribed mean curvature.   
We now use Theorem \ref{mu bubble} in the case $n=2$ to obtain Theorem \ref{size2}. 

\smallskip
Assume $(M,g,K)$ and $\S$ satisfy the hypotheses of Theorem \ref{size2}.  Choose points $p_-, p_+ \in \S$ such that ${\rm diam}(\S) = d(p_-,p_+)$.
Let $\g$ be a minimizing geodesic from $p_-$ to $p_+$ of length $\ell$.  Let 
$\S_{\e}$ be the compact surface-with-boundary obtained from $\S$ by removing from 
$\S$ the interiors of two small geodesic disks  $D^{\pm}_{\e}$ of radius $\e$, centered at $p_{\pm}$.  $\S_{\e}$ has  boundary 
consisting of two circles $C^{\pm}_{\e} = \d D^{\pm}_{\e}$.

Since, by Proposition  \ref{size1}, $\S$ is diffeomorphic to a $2$-sphere, 
$\S_{\e}$ is diffeomorphic to $S^1 \times [-1,1]$, and hence is a
$2$-dimensional overtorical Riemannian band. 
By applying Theorem \ref{mu bubble} in the case $n=2$, we obtain,
\beq
d(C^{-}_{\e},C^+_{\e}) \le \frac{2}{\sqrt {3}}\cdot \frac{\pi}{\sqrt{\Lambda_{\frac12}}} \,.
\eeq
Using the fact that $\g$ is a minimizing geodesic from $p_-$ to $p_+$, it follows that  
$$
d(C^{-}_{\e},C^+_{\e}) = \ell - 2\e = {\rm diam}(\S) - 2\e  \, ,
$$
and hence, 
\beq
{\rm diam}(\S) \le \frac{2}{\sqrt {3}}\cdot \frac{\pi}{\sqrt{\Lambda_{\frac12}}}   + 2\e 
\eeq
for all $\e$ sufficiently small. Thus to complete this proof of Theorem \ref{size2}, it suffices to observe that $\Lambda_{\frac12} \ge c$.

By Proposition \ref{keyfact} and the remark shortly after its statement,
$\l_1(\S_{\e}) \ge 0$, where $\l_1(\S_{\e})$ is the Dirichlet eigenvalue of the symmetrized operator 
\beq
\hat{L}(\phi)  = -\triangle \phi + \frac12 S_{\S} - \big(\mu + J(\nu) +
\frac12 |\chi_+|^2\big)\phi  \,.
\eeq

By Rayleigh's formula this implies 
\beq
\int_{\S_{\e}} |\D\psi|^2+ \Big[\frac12 S_{\S} - \big(\mu + J(\nu) + 
\frac12 |\chi_+|^2\big)\Big]\psi^2 \, dA   \ge 0
\eeq
for all $\psi \in C_0^{\infty}(\S_{\e})$.  Hence,
\begin{align}
\int_{\S_{\e}} |\D\psi|^2+ \frac12 S_{\S} \, dA  
&  \ge   \int_{\S_{\e}} \big(\mu + J(\nu) + \frac12 |\chi_+|^2\big)\psi^2 \, dA   \\
&  \ge c \int_{\S_{\e}} \psi^2 dA  \, ,
\end{align}
for all $\psi \in C_0^{\infty}(\S_{\e})$.  It now follows from Rayleigh's formula again that $\Lambda_{\frac12} \ge c$. \qed

\medskip
\noindent
\underline{\it Comments on higher dimensions.} 
The argument above to show that 
$\Lambda_{\frac12} \ge c$ holds for {\it any} stable MOTS $\S$ in any $n$-dimensional, $n \ge 3$, initial data set $(M,g,K)$ that satisfies $\mu - J(\nu) \ge c$.  

In \cite{HKKZ}, a Riemannian band $(N^n ,\partial_{\pm}N^n ,g)$ is referred to as a \textit{nonPSC-band} if $\partial_- N^n$ and $\partial_+ N^n$ are not separable by a smooth embedded hypersurface $V^{n-1}\subset N^n$ which admits a metric of positive scalar curvature.  As observed in \cite{HKKZ}, it follows from a classical result of Schoen and Yau that overtorical bands in dimensions $3\le n \le 7$ are nonPSC-bands.  Moreover, as is also pointed out, it follows from their proof of Theorem 1.3, that the theorem  holds more generally for nonPSC bands. Then given an $n$-dimensional Riemannian manifold $\Omega^n$, the authors  define the \textit{torical-radius} 
$\mathrm{Rad}_t(\Omega^n)$  to be the supremum of widths of all nonPSC-bands $(N^n ,\partial_{\pm}N^n ,g)$ that are isometrically immersed into $\Omega$.  Putting these facts together, one easily obtains the following. 

\begin{thm}\label{size3}
Let   $\S^{n-1}$ be a stable MOTS in an $n$-dimensional, $3 \le n \le 8$,  initial data set $(M^n,g,K)$.  Suppose there exists $c > 0$, such that  $\mu +J(\nu) \ge c$ on $\S^{n-1}$, where $\nu$ is the outward unit normal to $\S^{n-1}$.   Then the torical-radius satisfies,
\beq\label{ineq6}
{\rm Rad}_{t}(\S^{n-1}) \le  \pi\sqrt{\frac{2(n-1)}{n\, c}}  \,.
\eeq
\end{thm}

In \cite{HKKZ} this inequality is shown to hold, with $3 \le n \le 7$, for MOTS that arise from the cylindrical blow-up of Jang's equation, in connection with the authors's generalization of a result of Schoen and Yau \cite{SYsize} on the existence of black holes.   As the analysis of Schoen and Yau in \cite{SY2} showed for $3$-dimensional initial data sets (which was  extended to $n$-dimensions, $3 \le n \le 7$, by Eichmair \cite{Eich}), the Jang graph satisfies a stability property that is inherited by MOTS arising from the blow-up of Jang's equation, and implies that such MOTS are symmetric stable. The inequality~\eqref{ineq6} for such MOTS is a result of this analysis and Theorem 1.3 in \cite{HKKZ} (and its extension to nonPSC bands). The point we wish to make here is that \eqref{ineq6} holds for all stable MOTS up to dimension 7, regardless of how they arise.   Finally, we note that for $n \ge 3$ the torical-radius $\mathrm{Rad}_t(\Omega^n)$ does not directly relate to the diameter.

\medskip

\noindent
\textsc{Acknowledgements.} We are grateful to Marcus Khuri and Abra\~ao Mendes for helpful comments on earlier versions of this paper.   We also acknowledge research support from the Simons Foundation, under  Award No. 850541.


\providecommand{\bysame}{\leavevmode\hbox to3em{\hrulefill}\thinspace}
\providecommand{\MR}{\relax\ifhmode\unskip\space\fi MR }
\providecommand{\MRhref}[2]{%
  \href{http://www.ams.org/mathscinet-getitem?mr=#1}{#2}
}
\providecommand{\href}[2]{#2}

\end{document}